\documentclass[preprint2,11pt]{aastex}

\slugcomment{To Appear in Astrophysical Journal Letters}

\shorttitle{Roberts, Romani \& Johnston}
\shortauthors{Multiwavelength Studies of PSR J1420$-$6048}

\begin{document}


\title{Multiwavelength Studies of PSR J1420$-$6048, a Young Pulsar in the
Kookaburra}


\author{Mallory S.E. Roberts\altaffilmark{1}}
\affil{Dept. of Physics, McGill University, Montr\'eal, Qu\'e. H3A2T8 Canada}
\email{roberts@physics.mcgill.ca}

\author{Roger W. Romani}
\affil{Dept. of Physics, Stanford University, Stanford, CA 94025}
\email{rwr@astro.stanford.edu}

\and 

\author{Simon Johnston}
\affil{School of Physics, University of Sydney, NSW 2006, Australia}
\email{simonj@physics.usyd.edu.au}


\altaffiltext{1}{Center for Space Research, Massachusetts Institute of Technology, 
Cambridge, MA 02139}


\begin{abstract}
We present X-ray, radio, and infrared observations of
the 68~ms pulsar PSR J1420$-$6048 and its surrounding nebula,
a possible counterpart of the $\gamma-$ray source 
GeV J1417$-$6100/3EG J1420$-$6038.
Pulsed X-ray emission at the radio period is marginally detected by ASCA 
from a source embedded in the hard spectrum X-ray nebula 
AX J1420.1$-$6049.
At radio wavelengths, the pulsar is found to be strongly  
linearly and circularly polarized, and the polarization sweep is measured. 
A comparison of high resolution ATCA radio
imaging of the Kookaburra's upper wing (G313.6+0.3), which contains
the pulsar and the X-ray nebula, with infrared images 
suggests the radio emission is partly non-thermal.
\end{abstract}



\section{Introduction}

The X-ray source AX J1420.1$-$6049 is one of two
extended, hard spectrum sources whose positions are consistent with
the $\gamma-$ray source GeV J1417$-$6100/3EG J1420$-$6038 \citep{rr98,rrk01}.
Radio images of the region from the Molonglo Synthesis Telescope (MOST)
Galactic Plane Survey \citep{wcl94,g99} 
and the Australia Telescope Compact Array (ATCA) (Roberts et al. 1999, 
hereafter R99) 
reveal a complex of compact and extended radio sources. This complex, known as 
the Kookaburra, consists
of a shell with two ``wings" stretching to the northeast and the southwest. 
There is also a 
bright HII region and several compact extra-Galactic sources in the field. 
AX J1420.1$-$6049
is coincident with a slight enhancement (K3 of R99) in the 
northeast wing. 
Low resolution $IRAS$ 60~$\mu$m imaging of this region shows 
evidence of thermal emission from the shell  
but shows a lack of emission from the wings. However, the low resolution and 
artifacts from the HII region prevent strong constraints to be derived
from this image. 
The paucity of infrared emission along with the 
spectral index of the wing containing AX J1420.1$-$6049, although 
poorly constrained, hint at a non-thermal origin for the radio emission.

The 68.2~ms pulsar PSR J1420$-$6048 was recently discovered during 
the Parkes Multibeam
Survey at a position consistent with AX J1420.1$-$6049 \citep{d01}. 
Timing measurements
have determined a characteristic age of $\tau_c=1.3\times10^4$ yr, and a
very high inferred spin-down energy $\dot E=1.0\times 10^{37}$ ergs/s. 
This energy output is higher than all the known 
$\gamma$-ray pulsars except for the Crab,
and is similar to the recently discovered pulsar and pulsar
wind nebula (PWN) PSR J2229+6114, the
likely counterpart of the $\gamma-$ray source 3EG J2227+6122 \citep{h01}.
PSR J1420$-$6048 has been monitored since 1998 October, and was observed to 
glitch in late 2000. 
In this letter, we report on X-ray spectral and
timing analyses of a 1999 ASCA observation of AX J1420.1$-$6049, 
sensitive radio pulse profile and polarization measurements made with the Parkes 
radio telescope, high-resolution 20cm radio imaging with the 
ATCA, and high resolution 
infrared imaging at 60~$\mu$m and 8.3~$\mu$m. 
 
\section{ASCA Observations}

On Feb. 13-15, 1999,  ASCA re-observed GeV J1417$-$6100 centered on the
two X-ray nebulae.
Approximately 63~ks ($\sim22$ks with high-bit rate telemetry) of exposure 
was obtained with the Gas Imaging Spectrometers
(GIS2,GIS3) in standard PH mode with no timing bits.
Image and spectral analyses of the field have previously been reported in \citet{rrk01}, and
we use similar reduction and analysis procedures here.
Since the source fell near the edge of the
SIS detectors, we restrict our analysis to the GIS data.

AX J1420.1$-$6049 is seen by ASCA to be significantly
extended. 
However, there is a central concentration in AX J1420.1$-$6049 
whose centroided position is $\sim 40^{\prime \prime}$ from the radio position
of the pulsar (see below). This is larger than the nominal 90\% 
position uncertainty \citep{got00}, but is consistent with the 97\% contour;
we assume the peak of the X-ray emission is at the position of the pulsar
in the following analsysis.

\begin{figure}[h!]
\plotone{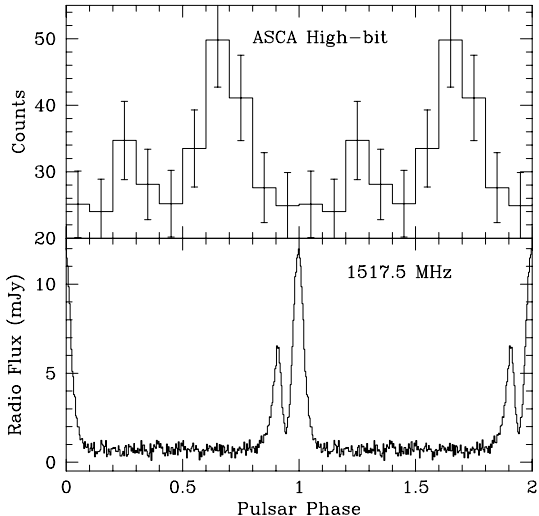}
\caption{\label{pulse}
The phase-aligned (absolute phase error $\sim 0.06$)
1-10 keV, ASCA GIS high bit-rate data (top) and 20~cm Parkes radio (bottom)
pulse profiles of PSR J1420$-$6038.}
\end{figure}

In general, timing solutions for young, glitching, pulsars such as PSR J1420$-$6048 
cannot be reliably extrapolated back to previous observation epochs. Therefore,
timing studies of the earlier ASCA dataset (1997 epoch) and the $EGRET$ data are problematic.
Fortunately, contemperaneous radio data is available for the 1999 ASCA observations 
\citep{d01}. These were observed in default observing mode 
which has a nominal time resolution of $62.5\mu s$. However, the reason
for such a large nominal
uncertainty in arrival time is a timing ambiguity which is only important
for sources with extremely high count rates ($\ga 10$ cps).
For low count rate sources such as PSR J1420$-$6048, the time assignment 
precision of the GIS is equal to the telemetry output period
($\sim 4$~ms for high bit-rate data, $31.25$~ms for medium bit-rate, 
Hirayama et al. 1996). 
Using the high bit-rate GIS data set, we extracted 1-10 keV photons from a 
6 pixel radius region ($\sim 1.5^{\prime}$, roughly  corresponding to the half-power 
radius of the ASCA PSF) centered on the position of the X-ray peak.
The small extraction region was chosen to minimize the photon contribution from the 
surrounding nebula. The resulting 314 photon arrival times were then 
corrected to the Solar System barycentre. Using
the radio period and period derivative to determine arrival time phases,
an H-test analysis was performed to determine the significance of any periodicity
\citep{d94} resulting in a test statistic value of 12.95. 
The formal chance probablity for a value this large 
is 0.0056, which we verified by examining the
H-test values of $\sim 12,000$ independent nearby periods 
on the same dataset.

The telemetry times used by ASCA are delayed from the true arrival time by
an offset of 8.5~ms $+ t_b$, where $t_b$ has an approximately uniform 
probability of 
being any value between $0$~s and $1/256$~s ($1/32$~s for the medium bit-rate
data) \citep{h96}. To extract the best folded light curve, 
we take into account this delay as well as the arbitrary 
placement of bin boundaries by assigning a fractional value to adjacent
bins according to the probability of a photon belonging in each.  
The result is shown in Figure~\ref{pulse}. The absolute phasing 
of this method was verified with an offset observation of the
Crab pulsar taken in the same mode and folded using the Jodrell Bank ephemeris. 
For PSR J1420-6048, the X-ray pulse ranges in phase from $\sim0.60-0.75$ 
referenced to the radio pulse maximum. 
A similar analysis with the medium bit-rate data  resulted in
a profile consistent in amplitude and phase with a highly smoothed version 
of the high bit-rate data. The formal H-test chance probability of the
medium bit-rate data is only 0.14.

\begin{figure}[h!]
\plotone{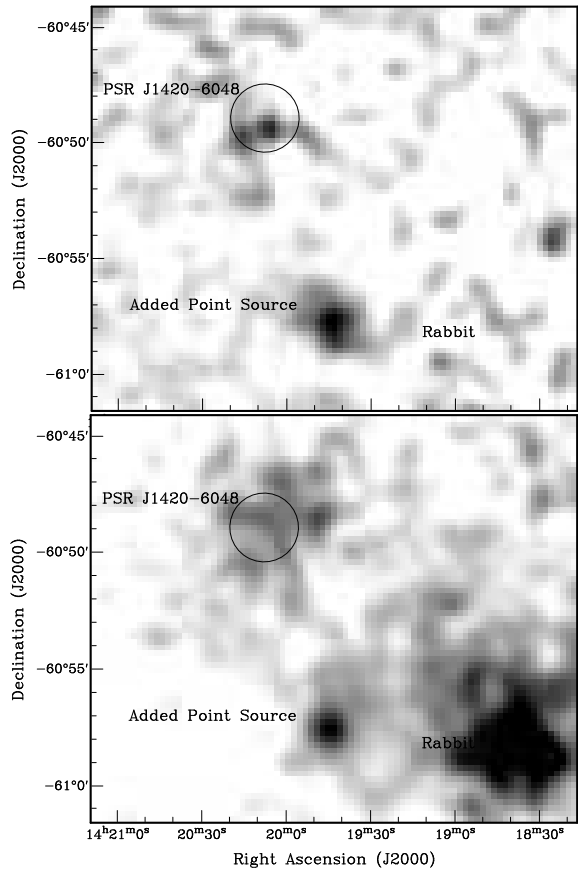}
\caption{\label{onoff} Pulsed-only (top) and off-pulse (bottom) images of
pulsar region made
from high bit rate data only (see text) smoothed with a $\sigma=1.7$ pixel
Gaussian. A
calibration point source has been added for comparison. The circle
represents the extraction region for the timing analysis.
}
\end{figure}

The above phase window was used to make on-pulse and off-pulse 1-10 keV images
from the high bit-rate data. 
The off-pulse image was scaled and subtracted from the on-pulse image to
create a pulsed-only image, which was then exposure corrected and smoothed with
a 1.7 pixel Gaussian (roughly the width of the GIS PSF peak).
A particle background image made from the night earth calibration files
was subtracted from the off-pulse image, which was also exposure
corrected and smoothed. The resulting pulsed-only and off-pulse images
are presented in Figure~\ref{onoff}. A radial profile comparison of these 
images to point-source calibration images for nearby positions on the detector,
(one such is added as a reference to the images in Figure~\ref{onoff})
show the pulsed-only profile to be consistent with
that of the calibration source while the off-pulse profile is virtually flat
out to $r\sim1^{\prime}$ and then slowly falls off.  This provides additional
support to the pulse detection.

The overall X-ray flux for the entire data set of high and medium bit-rate data 
is adequately fit by a power-law with $\Gamma =1.6\pm 0.4$
with significant absorption ($N_H=(2.2\pm 0.7)\times 10^{22} {\rm cm^{-2}}$)
and a 2-10~keV flux of $4.7 \times 10^{-12} {\rm erg/cm^2s^{-1}}$. Weak evidence
for an Fe K line (EW$=340\pm260$~eV) suggests that some thermal emission
may also be present in the region.
In order to estimate the magnitude of the apparent pulsed flux, a 
$3^{\prime}$ 
extraction radius was used to create on-pulse and off-pulse spectra from the
high bit-rate data using a slightly larger phase range of 0.575-0.775 to
maximize inclusion of the pulsed emission. This extraction region
is roughly twice the point spread function full width half max, but
is
smaller than the nominal $6^{\prime}$ in order to limit noise due to photons
from the surrounding nebula.
Using the off-pulse spectrum as a background, and fixing the 
$N_H$ at the value for AX J1420.1$-$6049 as a whole,
the resulting spectrum is adequately fit by either a power law or blackbody, 
with a 1-10 keV flux of roughly $5\times 10^{-13}$ ergs ${\rm cm}^{-2}{\rm s}^{-1}$.

\section{Parkes Radio Timing Observations}

PSR J1420$-$6048 was observed on May 20, 2001 with the
Parkes 64m radio telescope as part of a program of high time-resolution 
observations of Vela-like pulsars \citep{jr02}. 
The central feed of the 13 beam, 21~cm
multi-beam system was used.  The signal was split in two and sent to
two different back-ends. One was 
a dual polarization, 512 channel filterbank receiver with a 
central observing frequency of 1517.5~MHz and a 256~MHz bandwidth, where
the two polarizations were summed, 
fast sampled, one-bit digitized and written to storage tape  
for off-line de-dispersion
and pulse folding. This was used for the high-resolution profile. The second
back-end was a 128 MHz bandwidth correlator with somewhat lower frequency
resolution, but which retained the full polarization information and was
used for the polarization analysis.

A 4 hour integration on PSR J1420$-$6048
folded at the contemporary radio ephemeris produced the high resolution
pulse profile in Figure 1. The pulsar has a broad profile ($\sim 83^{\circ}$
= 16~ms at 10\% of the peak flux) with two well separated components.
We have set radio phase 0 (the maximum of the
second peak) to the phase of the radio pulse at the (pre-glitch) ephemeris
used to fold the X-ray data.

\begin{figure}[h!]
\plotone{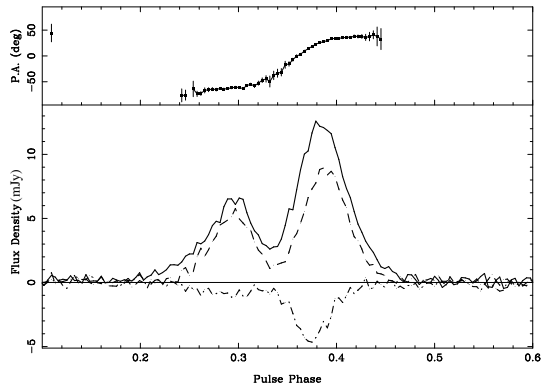}
\caption{\label{pks}Pulse polarization for PSR J1420-6048. Below: The total
intensity
(full line) linearly polarized  intensity (long dash line) and circularly
polarized intensity (dot-dashed line). The $1\sigma$ flux errors are
shown by the
error bars to the upper left.  Above: The linear polarization position angle
sweep with associated error bars.}
\end{figure}

Figure~\ref{pks} shows the polarization profiles of the radio peaks
at $\sim 2$\,ms effective time resolution.
The pulse shows a high fractional linear polarization ($\sim 65\%$), with
a smooth position angle swing. High linear polarization
appears to correlate with high spindown luminosity ${\dot E}$
\citep{cmk01}
and PSR J1420$-$6048 supports this trend.  A simple \citet{rc69}
rotating vector model fit to the position angle produces an 
adequate fit to the sweep for small impact parameters ($\beta \sim 0.5^\circ$)
and small magnetic inclinations $\alpha \la 35^\circ$, although the
latter angle is relatively poorly determined. 
We have also extracted the rotation measure 
$({\rm RM}= -106 \pm 18\, {\rm rad \, m^{-2}})$, which together with the 
previously 
measured dispersion measure gives a mean line-of-sight magnetic field strength of
$\sim 0.4 \,\mu {\rm G}$, pointed away from the observer. 
The pulsar also shows 
very strong (right handed) circular polarization, particularly in the 
second peak where the circular polarization reaches $\sim 35\%$.  This
value is among the highest recorded (Han et al. 1998), certainly the highest
among pulsars classified as `conal doubles'. The 
right-handed circular polarization and increasing position angle swing 
follow the correlation between handedness and sense of position
angle change found by Han et al. (1998) for cone-dominated pulsars.
With no reversal of sign, the phase averaged circular polarization
fraction is quite large at 23\%.

\section{Radio and Infrared Imaging}

\begin{figure}[h!]
\plotone{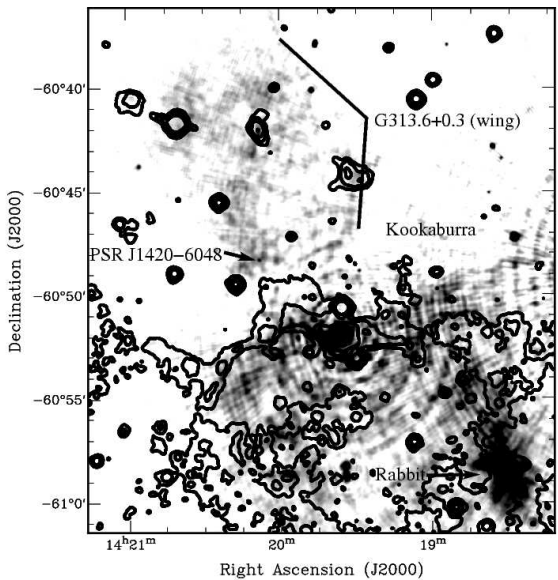}
\caption{\label{hires}
High resolution 20~cm image (greyscale) shows the structure in the
upper ``wing" G313.6+0.3 and the pulsar.
The $MSX$ $8.3\mu$ image (contours) shows
several compact sources within G313.6+0.3 but no extended emission.}
\end{figure}

R99 presented results of radio continuum observations at  20~cm and 13~cm
of the
Kookaburra complex with the ATCA in three antenna configurations
(1.5A, 0.375, and 0.75A configurations on 11 Jan., 28 Mar., and 22 Apr., respectively.)
Three additional 20~cm continuum observations have since been performed
(6D, 0.375, and 0.75C configurations on 20 Aug. 1999, 13 Jan. 2001, 
and 03 Mar. 2001). 
All these data were reduced and analysed
with the Miriad software package \citep{sk98}.
A high resolution 
($10.2^{\prime \prime}\times 9.9^{\prime \prime}$ beam, 1375 MHz mean 
frequency) image was produced by
combining all of the 20~cm data. Figure~\ref{hires} shows
the upper wing of the Kookaburra 
which we name G313.6+0.3 (K2 of R99). 
The pulsar is clearly seen as a point source near the southeast corner 
at R.A.(J2000)=14h 20m 8.23(6)s,
Dec.(J2000)=$-60^{\circ}$ $48^{\prime}$ $16.8(2)^{\prime \prime}$,
consistent with the timing position of PSR J1420$-$6048
\citep{d01}. Stokes V parameter imaging finds a circular polarization
fraction consistent with the time averaged value from the Parkes data.
Using the dual-frequency 1998 data and only the longest ($\sim 6$~km) baselines
to resolve out the continuum flux, 
we obtain pulse average fluxes of $0.7\pm0.1$ mJy (20cm) and
$0.7\pm0.1$mJy (13cm) for a net spectral index $\alpha \sim 0$.

\citet{wg96} have used a low ratio of infrared
to radio flux as an indicator of non-thermal emission in order to 
identify supernova remnant candidates. 
We have obtained {\it IRAS} Hi-res $60~\mu$m and
{\it MSX} $8.3~\mu$m images from the NASA/IPAC Infrared Science Archive, 
and compared them to the radio maps. The shell is clearly seen in infrared,
but there is no excess emission associated with 
G313.6+0.3 (Figure~\ref{hires} contours) in either image. 

\section{Discussion}

\citet{yr97} noticed a plausible association between the $\gamma-$ray
source and a complex
of young star forming regions and SNR at a distance $d\la 2$~kpc. 
However, the pulsar dispersion measure (DM=360 pc ${\rm cm}^{-3}$, 
D'Amico et al. 2001) 
indicates a much larger 
distance $d\approx 7.7\pm 1.1$~kpc \citep{tc93}. The measured X-ray
absorption of $N_H\sim2\times 10^{22} {\rm cm}^{-2}$ 
is consistent with the total 
Galactic $N_H$ towards that direction estimated from the FTOOLS $nh$ tool 
derived  from  the  HI  map by \citet{dl90}. However,  
the prevalence of HII regions and dense
clouds in this direction may mean the DM and $N_H$ are dominated by local 
sources in the Carina arm, implying a closer distance. 

We can compare the X-ray and radio fluxes of our sources with other PSR/PWN.
Saito (1997) fits an empirical relation for pulsed
ASCA fluxes of $L_{pulse} (2-10~keV) = 10^{34}{\dot E_{38}}^{3/2}$
erg/s, where ${\dot E_{38}}$ is the spin-down power in 
units of is $10^{38}$erg/s.
This matches our observed
pulsed flux at $d \approx 2$~kpc. 

The radio flux can also be compared to known PWN. R99 suggested that the K3 
site,
a $\sim 20$ mJy enhancement above the general flux in G313.6+0.3, was
the radio counterpart to AX J1420.1-6049.
Assuming a typical radio PWN spectral
index $\alpha = -0.3$, \citet{gs00} note the fraction $\epsilon$ of
the spin-down energy which goes into radio emission is related to the
20~cm flux density by 
$S_{20cm}\approx 3.3\times 10^6\epsilon\dot E_{37}/(d/8{\rm kpc})^2$ mJy. 
This suggests an efficiency $\epsilon \sim 6\times 10^{-6}$
at a distance of 8 kpc which is rather low compared to most other
detected radio PWN \citep{fs97,gs00}, and would be even lower if the true 
distance is closer. 
However, the entire wing has a flux of $S_{20cm}\sim 1$ Jy, the nature of which 
is uncertain.
It is possible that a substantial fraction of the flux is due to the pulsar 
wind. 
A PWN flux of $\sim 200$~mJy would make for a fairly typical value of
$\epsilon \sim 6\times 10^{-5}$ at 8~kpc, and reasonable values at closer distances. 
The implied size of the nebula at 8~kpc, $\sim 20\times 30$ pc,
would then be unusually large. 

The morphology of G313.6+0.3 is curious, being a rectangular plateau
of radio emission with a question mark shaped enhancement starting at the
pulsar and reaching up to nearly the top, reminiscent of a
trailing bow-shock.
The lack of IR emission, along with the apparent excess of polarized
flux seen by R99 in portions of the wing and their estimate of $\alpha=-0.2\pm0.2$ for the spectral index, suggest the entire wing may be a supernova remnant.
However, the large scatter in the T-T plot of R99 makes their spectral index
value problematic.
The expanded data set confirms the presence of linear polarizaton near the
top of the enhancement, but seems to suggest that G313.6+0.3 has a mixture 
of thermal and non-thermal spectral components.

Given the rarity of such high ${\dot E}$ pulsars, it is natural
to assume a connection with GeV J1417$-$6100/3EG J1420$-$6038. The average
$E\ge 100$~MeV flux for this source is $4 \times 10^{-7} \gamma/{\rm cm^2/s}$.
Models of $\gamma$-ray pulse emission (e.g. Romani 1996) allow one to
predict the typical $\gamma$-ray flux for pulsar spin parameters and
the magnetic inclination angle $\alpha$. For PSR J1420$-$6048 ($\tau_c = 
1.3 \times 10^4$yr, surface magnetic field $B = 2.4 \times 10^{12}$G), 
an assumed 1 steradian beam, and an efficiency of $10^{-3}$
(somewhat less than Vela) the observed $E\ge 100$MeV flux
is expected for $\alpha =50^\circ$ if the distance were as close as
2~kpc.  At a distance of 8~kpc, 
the inferred efficiency would need to be significantly higher if the
$\gamma-$ray emission is 100\% pulsed. If $\alpha$ is truely less than 
$35^{\circ}$, the expected $\gamma$-ray 
efficiency is quite small and in an outer gap picture one would not
expect the $\gamma$-ray beam to cross the line of sight for small
$\beta$. In this case the $\gamma$-ray emission could have an unpulsed,
pulsar wind origin.

It is important to note that
the $\gamma$-ray flux appears variable. A significant fraction
of the emission may therefore
be associated with a pulsar wind shock, where instabilities
might produce luminosity fluctuations on $\sim$month time scales.
Similar variability is seen in other unidentified {\it EGRET} 
sources containing PWN
\citep{rgr01}. 
A second pulsar in the nearby Rabbit PWN candidate, unseen in the radio,
might also contribute to the $\gamma-$ray flux.

\acknowledgments

We thank Bryan Gaensler, Vicky Kaspi, and the referee  Eric Gotthelf for 
useful comments. We also thank
Froney Crawford for the 20 Aug. 1999 ATCA data.
The Australia Telescope is funded by the commonwealth of Australia for
operation as a National Facility managed by the CSIRO. MSER is a 
Quebec Merit Fellow. This work was 
supported in part by NASA grants NAG5-3263 and a Cottrell Scholars
grant from the Research Corporation.

\end{document}